\documentclass[prd,aps,amssymb,nofootinbib,showpacs,preprint]{revtex4}
\usepackage{amsmath,amsfonts}

\usepackage{hyperref}

\newcommand{\beq}{\begin{equation}}
\newcommand{\eeq}{\end{equation}}
\newcommand{\bear}{\begin{eqnarray}}
\newcommand{\ear}{\end{eqnarray}}
\newcommand{\earn}{\nonumber \end{eqnarray}}
\newcommand{\dst}{\displaystyle}
\newcommand{\nn}{\nonumber \\}
\newcommand{\Ref}[1]{(\ref{#1})}

\begin{document}
%--------------------------------------------------------------------

\title{Vacuum polarization of a quantized scalar field in the thermal state
 in a long throat}

\author{Arkady A. Popov\thanks{Email address: apopov@ksu.ru}}

\address{Kazan Federal University, 18 Kremlyovskaya St., Kazan 420008, Russia}

\begin{abstract}

Vacuum polarization of scalar fields in the background of a long throat  is investigated. 
The field is assumed to be both massive or massless, with arbitrary coupling to the scalar 
curvature, and in a thermal state at an arbitrary temperature. Analytical approximation
for $\left< \varphi^2 \right>_{ren}$ is obtained.
\end{abstract}

\maketitle
%--------------------------------------------------------------------

\section{Introduction}

%--------------------------------------------------------------------

The study of vacuum polarization effects in strong gravitational fields is
a pertinent issue since such effects may play a role in the cosmological
scenario and in the construction of a self-consistent model of black hole
evaporation. These effects may be taken into account by solving the semiclassical
backreaction equations,
     \beq \label{1}
     G^{\mu}_{\nu}=8 \pi \langle T^{\mu}_{\nu} \rangle_{ren},
     \eeq
where $\langle T^{\mu}_{\nu} \rangle_{ren}$ is the expectation value of the stress-energy
tensor operator for the quantized fields. Possible vacuum fluctuations of quantized fields
can create wormholes \cite{MT,Sushkov92,HPS,PopovCQG}.

The main difficulty in the theory of semiclassical gravity is that
the effects of the quantized gravitational field are ignored. The
popular solution to this problem is to justify ignoring the
gravitational contribution by working in the limit of a large
number of fields, in which the gravitational contribution is
negligible. Another problem is that the vacuum polarization
effects are determined by the topological and geometrical
properties of spacetime as a whole or by the choice of quantum
state in which the expectation values are taken. It means that
calculation of the functional dependence of $\langle T^{\mu}_{\nu}
\rangle_{ren}$ on the metric tensor in an arbitrary spacetime
presents formidable difficulty. Only in some spacetimes with high
degrees of symmetry for the conformally invariant fields $\langle
T_{\mu \nu} \rangle_{ren}$ can be computed and equations \Ref{1}
can be solved exactly \cite{KSS,KSS2,KSS3,KSS4,KSS5}.

Numerical computations of $\langle T^{\mu}_{\nu} \rangle_{ren}$ are usually extremely intensive
\cite{HC,C,C2,C3,C4,C5,AHS,BBK}. In some cases $\langle T_{\mu \nu} \rangle_{ren}$ is determined
by the local properties of a spacetime and it is possible to calculate the functional
dependence of the renormalized expression for the vacuum
expectation value of the stress-energy tensor operator of the
quantized fields on the metric tensor approximately (too long). One of the
most widely known examples of such a situation is the case of a
very massive field. In this case $\langle T^{\mu}_{\nu} \rangle_{ren}$ can be expanded in terms
of powers of the small parameter
      \beq \label{ml}
      \frac1{m l} \ll 1,
      \eeq
where $m$ is the mass of the quantized field and $l$ is the
characteristic scale of the spacetime curvature \cite{MatK,MatK2,MatK3,MatK4,MatK5,MatK6,MatK7}.

Approximate calculations for conformally coupled massless fields have
also been made. For ${ \langle T^{\mu}_{\nu}\rangle }$ in static
Einstein spacetimes ($R_{\mu \nu}=\Lambda g_{\mu \nu}$) these
include the approximations of Page, Brown, and Ottewill
\cite{Page,BO,BOP}. These results have been generalized to
arbitrary static spacetimes by Zannias \cite{Z}. A different
approach to the derivation of approximate expressions for
${\langle \varphi^2 \rangle}$ and ${\langle T^{\mu}_{\nu}
\rangle}$ for conformally coupled massless fields in static
spacetimes has been proposed by Frolov and Zel'nikov \cite{FZ}.
Their calculations were based primarily on geometric arguments and
the common properties of the stress-energy tensor rather than on a
field theory. Using the methods of quantum field theory the
expressions for ${\langle \varphi^2 \rangle}$ and ${\langle
T^{\mu}_{\nu} \rangle}$ of a scalar field in static spherically
symmetric asymptotically flat spacetimes have been obtained by
Anderson, Hiscock, and Samuel \cite{AHS}. They assumed that the
field is massive or massless with an arbitrary coupling $\xi$ to
the scalar curvature and in a zero temperature quantum state or a
nonzero temperature thermal state. The result was presented as a
sum of two parts, numerical and analytical:
      \beq
      \langle T^{\mu}_{\nu} \rangle_{ren}=\langle T^{\mu}_{\nu}
      \rangle_{numeric}+\langle T^{\mu}_{\nu} \rangle_{analytic}.
      \eeq
The analytical part of their expression is conserved. This has a
trace equal to the trace anomaly for the conformally invariant
field. For these reasons they proposed to use $\langle
T^{\mu}_{\nu} \rangle_{analytic}$ directly as an approximation for
$\langle T^{\mu}_{\nu} \rangle_{ren}$. An analogous result has
been obtained by Groves, Anderson, and Carlson \cite{GAC}  in the
case of a massless spin $\frac{1}{2}$ field in a general static
spherically symmetric spacetime.

Let us stress that the single parameter of length dimensionality in problem (\ref{1}) is
the Planck length ${l_{ \mbox{\tiny \sl Pl}}}$. This implies that
the characteristic scale $l$ of the spacetime curvature (which
corresponds to the solution of equations (\ref{1})) can differ from
${l_{ \mbox{\tiny \sl Pl}}}$ only if there is a large
dimensionless parameter. As an example of such a parameter one can
consider a number of fields the polarization of which is a source
of spacetime curvature \footnote{Here and below it is
assumed, of course, that the characteristic scale of change of the
background gravitational field is sufficiently greater than ${l_{
\mbox{\tiny \sl Pl}}}$ so that the very notion of a classical
spacetime still has some meaning.}.
In the case of massive field, the existence of an
additional parameter $1/m$ does not
increase the characteristic scale of the spacetime curvature $l$ which
is described by the solution of equations (\ref{1}) \footnote{The
characteristic scale of the components $G^{\mu}_{\nu}$ on the
left-hand side of equations (\ref{1}) is $1/l^2$, on the
right-hand side - ${l_{ \mbox{\tiny \sl Pl}}}^2/(m^2 l^6)$}.
For the massless quantized fields such a parameter can be the coupling constants of field
to the curvature of spacetime \cite{PopovCQG}.
Another possibility of introducing an additional parameter in the problem (\ref{1}) is
to consider the non-zero temperature of quantum state for the quantized field.
It is known (see, e.g., \cite{NF}) that in the high-temperature limit (when $T \gg 1/l, T$ being
a temperature of thermal state) $\langle T^{\mu}_{\nu} \rangle$ for such a thermal state
is proportional to the fourth power of the temperature $T$.

In this paper, approximate expression for ${\langle \varphi^2 \rangle}_{ren}$  of a quantized
scalar field in the background of a long throat is derived. The field is assumed to be both
massless or massive with a coupling $\xi$ to the scalar curvature and in a thermal
state at an arbitrary temperature.

In Sec. II the expression for the Euclidian Green's function of a scalar field
in a general static spherically symmetric spacetime is derived.
In Sec. III the WKB approximation for ${\langle \varphi^2 \rangle_{unren}}$ is obtained
and the renormalization procedure is described.
The results are summarized in Sec. IV.
Here we give also an example of the evaluation ${\langle \varphi^2 \rangle_{ren}}$ near ultraextremal
horizon.
In the Appendixes some technical results are derived and long expressions are displayed.

The units $\hbar=c=G=k_B=1$ are used throughout the paper.

%%--------------------------------------------------------------------

\section{An unrenormalized expression for $\left< \varphi^2 \right>$ }

%%--------------------------------------------------------------------
The metric of the static spherically symmetric spacetime is analytically continued into Euclidean form
  \beq\label{metric}
  ds^2= f d\tau^2+d\rho^2+r^2(d\theta^2+\sin^2\theta\, d\varphi^2),
  \eeq
where $f$ and $r$ are functions of $\rho$, and $\tau$ is the Euclidean time ($\tau = it,$ where $t$
is the coordinate corresponding to the timelike Killing vector, which always exists
in static spacetime).

In this paper, the point-splitting method is employed for the regularization of ultraviolet divergences.
The expectation value $\langle\phi^2(x)\rangle$ is found  from the coincidence limit of the Euclidean
Green’s function $G_E({x},{x})\equiv \lim_{\tilde x \rightarrow x} G_E({x},{\tilde x})$.
For a scalar field in a spacetime given by Eq.\ (\ref{metric}),
$G_E({x},{\tilde x})$ satisfies
\beq
\left[ \square_x -m^2-\xi R(x) \right] G_E(x,\tilde x)=-\frac{\delta^4(x,\tilde x)}{\sqrt{g(x)}},
\eeq
where $\square_{x}$ is the Laplace-Beltrami operator of the Euclidean metric corresponding
to Eq.\ (\ref{metric}), $m$ is the mass of scalar field $\phi$ with coupling $\xi$ to
the scalar curvature $R$ of spacetime.

The representation for the Euclidean Green's function
$G_E(x,\tilde x)$ of a scalar field in a static spherically
symmetric spacetime used by Anderson \emph {et al.} \cite{AHS} is
the following:  %\omega = 2 \pi n T
      \bear \label{GE}
      G_E(x;\tilde x)&=&\frac{T}{4 \pi}
       \sum \limits_{l=0}^{\infty}
      (2l+1) P_l (\cos \gamma ) \ C_{0 l} \ p_{0 l}(\rho_<)
      \ q_{0 l}(\rho_>) \nn &&
      +\frac{T}{2 \pi}\sum \limits_{n=1}^{\infty } \cos [\frac{}{}2\pi n T (\tau  - \tilde \tau)]  \sum \limits_{l=0}^{\infty}
      (2l+1) P_l (\cos \gamma ) \ C_{n l} \ p_{n l}(\rho_<)
      \ q_{n l}(\rho_>),
      \ear
where $P_l$ is a Legendre polynomial, $T$ is the temperature of the field,
$\cos \gamma \equiv \cos \theta \cos \tilde \theta+\sin \theta \sin \tilde
\theta \cos (\varphi -\tilde \varphi)$, $C_{n l}$ is a normalization
constant, $\rho_<$ and $\rho_>$ represent the lesser and greater
of $\rho$ and $\tilde \rho$, respectively, and the modes
$p_{n l}(\rho)$ and $q_{n l}(\rho)$ satisfy the equation
   \bear\label{modeeqn} \left\{ {
   \frac{d^2}{d\rho^2}+\left[\frac{1}{2f}\frac{df}{d\rho}
   +\frac{1}{r^2}\frac{dr^2}{d\rho}\right]\frac{d}{d\rho}
   -\left[\frac{(2 \pi n T)^2}{f}+\frac{l(l+1)}{r^2}+m^2
   +\xi R\right]}\right\}\left\{
   {\begin{array}{l}p_{n l}\\
   q_{n l}\end{array}}\right\}=0.
   \ear
Normalization of $p_{n l}$ and $q_{n l}$ is given by the Wronskian condition
        \beq\label{wronskian}
        C_{n l}\left[p_{n l}\frac{dq_{n l}}{d\rho}-
        q_{n l}\frac{dp_{n l}}{d\rho}\right]=\frac{-1}{r^2f^{1/2}}.
        \eeq
Above, it is assumed that the field is in a nonzero temperature
vacuum state defined with respect to the timelike Killing vector.

By the change of variables
        \beq\label{modes}
        \begin{array}{l}
        \displaystyle
        p_{n l}=\frac1{\sqrt {2 r^2 W}}
        \exp \left\{ \int^\rho W f^{-1/2} d\rho \right\}, \\
        \\
        \displaystyle
        q_{n l}=\frac1{\sqrt{2 r^2 W}}
        \exp\left\{- \int^\rho W f^{-1/2} d\rho \right\},
        \end{array}
        \eeq
one sees that the Wronskian condition (\ref{wronskian}) is satisfied by
\beq C_{n l}=1 \eeq
and the mode equation (\ref{modeeqn})
gives the following equation for $W(\rho)$:
     \beq \label{W2}
     W^2=(2 \pi n T)^2+\frac{f}{r^2} \left[\frac{}{} l(l+1) +2 \xi +m^2 r^2 \right]
     +{ \frac{f'}{8}\frac{(W^2)'}{W^2} +\frac{f}{4}\frac{(W^2)''}{W^2} -\frac{5f}{16}\frac{(W^2)'^2}{W^4} }+V,
     \eeq
where
     \bear
     V&=& f\left( { \frac{(r^2)''}{2r^2}+\frac{f'(r^2)'}{4fr^2}
      -\frac{(r^2)'^2}{4r^4}} \right) % \nn  &&
    + \xi f \left( {-\frac{f''}{f}-2\frac{(r^2)''}{r^2}
    +\frac{f'^2}{2f^2} +\frac{(r^2)'^2}{2r^4}-\frac{f'(r^2)'}{fr^2}
     } \right).
     \ear
The prime denotes a derivative with respect to $\rho$.

Now one can write expression for $\left< \varphi^2 (x, \tilde x) \right>_{unren} = G_E(x,\tilde x)$
using expressions (\ref{GE}),(\ref{modes}) and then suppose
$\rho=\tilde \rho, \ \theta=\tilde \theta, \ \phi=\tilde \phi$ ($P_l(1)=1$).
The superficial divergences in the sums over $l$ that appear in
this case can be removed as in Refs. \cite{HC,AHS,CH2,H,A}:
     \bear\label{B11}
     \left< \varphi^2 \right>_{unren}&=&\frac{T}{4 \pi r \sqrt{f}}
     \sum \limits_{l=0}^{\infty}
     \left[ \sqrt{\frac{f}{r^2}} \frac{(l+1/2)}{W_{|n=0}}-1 \right] \nn
     &&+ \frac{T}{2 \pi r \sqrt{f}} \sum_{n=1}^{\infty} \cos(u_n \varepsilon)
     \sum \limits_{l=0}^{\infty}
     \left[ \sqrt{\frac{f}{r^2}} \frac{(l+1/2)}{W}-1 \right],
      \ear
      \beq \label{u}
      \varepsilon=\sqrt{\frac{f}{r^2}}(\tau-\tilde \tau), \quad
      u_n=2 \pi n T \sqrt{\frac{r^2}{f}}.
      \eeq

%%--------------------------------------------------------------------
\section{WKB approximation}
%%--------------------------------------------------------------------

Equation (\ref{W2}) can be solved iteratively when
the metric functions $f(\rho)$ and $r^2(\rho)$ are varying slowly,
that is,
     \beq \label{lwkb}
     \varepsilon_{\mbox{\tiny \sl WKB}}=L_{\star} /L \ll 1,
     \eeq
where
      \beq \label{Lst}
      L_{\star}(\rho) =\frac{r(\rho)}{\sqrt{|2\xi+m^{2}r^{2}(\rho)|}},
      \eeq
and $L$ is a characteristic scale of variation of the metric
functions:
       \beq \label{Lm}
       \frac{1}{L(\rho)}= \max \left \{ \left| \frac{r'}{r}  \right|, \
       \left| \frac{f'}{f}  \right|, \
        \left| \frac{r'}{r} \sqrt{\left|\xi \right|} \right|, \
        \left| \frac{f'}{f} \sqrt{\left|\xi \right|} \right|, \
       \left| \frac{r''}{r}  \right|^{1/2}, \
       \left| \frac{f''}{f}  \right|^{1/2}, \
       \dots  \right \} .
       \eeq
We shall call the region of spacetime where the metric functions $f(\rho)$ and
$r^2(\rho)$ is slowly varying \textsl{the long throat}. This type of region exists,
for example, in the neighborhood of the ultraextremal  horizon \cite{PopZas:2007}.

The zeroth-order WKB solution of Eq. (\ref{W2}) corresponds
to neglecting terms with derivatives in this equation
       \beq
       W^2=U_0,
       \eeq
where
        \bear
        U_0&=&(2 \pi n T)^2+\frac{f}{r^2}\left(l+\frac12\right)^2
        +\frac{f}{r^2}\mu^2, \nn
       \mu^2&=&m^2r^2+2\xi-\frac14.
        \ear
Let us stress that $U_0$ is the exact solution of Eq. (\ref{W2}) in
a spacetime with metric
$ds^2=f_0d\tau^2+d\rho^2+r_0^2(d\theta^2+\sin^2\theta\, d\varphi^2)$,
where $f_0$ and $r_0$ are constants.

Below as in Ref. \cite{PS}  it is assumed that
       \bear
       \mu^2>0.
       \ear
The forth-order solution is
        \beq\label{wkbsol}
        W^2=U_{0}+U_{2}+U_{4}
        \eeq
with
        \beq
        U_{2}=V+\frac{f'}{8} \frac{U_{0}{}'}{U_{0}}
        +\frac{f}{4} \frac{U_{0}{}''}{U_{0}}
        -\frac{5f}{16} \frac{{U_{0}{}'}^2}{U_{0}{}^2}
        \eeq
and
        \bear
        U_{4}&=&\frac{f'}{8}\left( \frac{U_{2}{}'}{U_{0}}-\frac{U_{0}{}' U_{2}}{U_{0}{}^2}\right)
        +\frac{f}{4} \left(\frac{U_{2}{}''}{U_{0}} - \frac{U_{0}{}'' U_{2}}{U_{0}{}^2} \right)
        -\frac{5f}{8} \left( \frac{U_{0}{}' U_{2}{}'}{U_{0}{}^2} - \frac{{U_{0}{}'}^2 U_{2}}{U_{0}{}^3} \right)
       \nn &=& \frac{1}{U_0}\left[\frac{f' V'}{8}
      +\frac{f V''}{4}\right]
      +\frac{1}{{U_0}^2}
      \left[\left(\frac{f' f''}{64}-\frac{5fV'}{8}+\frac{f f'''}{32}
      -\frac{f' V}{8}\right){U_0}'
      \right. \nn && \left.+\left(\frac{f f''}{8}
      +\frac{3 f'^2}{64}-\frac{f V}{4}\right)
      {U_0}''
      + \frac{3ff'}{16}{U_0}'''
      + \frac{f^2}{16}{U_0}''''\right] \nn
      &&+\frac{1}{{U_0}^3}\left[
      \left(-\frac{9 f'^2}{128}-\frac{7ff''}{32}
      +\frac{5fV}{8}\right){{U_0}'}^2
      -\frac{15ff'}{16}{U_0}' {U_0}''
      -\frac{9f^2}{32}{{U_0}''}^2
      \right. \nn && \left.
      -\frac{7f^2}{16}{U_0}' {U_0}'''
      \right]
      +\frac{1}{{U_0}^4}\left[
      \frac{27ff'}{32}{{U_0}'}^3
      %\right. \nn && \left.
      +\frac{27f^2}{16} {{U_0}'}^2{U_0}''
      \right]
      -\frac{1}{{U_0}^{5}}\left[
      \frac{135f^2}{128}{{U_0}'}^4
      \right].
      \ear

\noindent
The approximate (of fourth WKB order) expression for $\left< \varphi^2 \right>_{unren}$ is obtained by substituting the WKB expansion
of $W^2$ [Eq. (\ref{wkbsol})] into Eq. (\ref{B11})):

\bear\label{B1_}
&&\left< \varphi^2 \right>_{unren}=\frac{1}{4\pi^2} \left\{
{\frac {1}{r^2}} {\bf S^0_0(\varepsilon,\mu)}
-{\frac {V}{2f}} {\bf S^0_1(\varepsilon,\mu)}
-\frac{r^2}{16 f^2}{\left[f'\left( \frac{f}{r^2} \right)'
+2 f\left( \frac{f}{r^2} \right)''\right]} {\bf S^1_2(\varepsilon,\mu)} \right. \nn &&
-\frac{r^2}{16f^2}
\left[-6 V^2+2f\left( \mu^2 \frac{f}{r^2} \right)''+f'V'
+f'\left( \mu^2 \frac{f}{r^2} \right)'+2fV''\right] {\bf S^0_2(\varepsilon,\mu)}
\nonumber
\ear
\bear
&&
+{\frac {5r^4}{32f^2}}\left( \frac{f}{r^2} \right)'^{2} {\bf S^2_3(\varepsilon,\mu)}
-\frac{r^4}{128 f^3}
\left [f'f''\left( \frac{f}{r^2} \right)'
+8ff''\left( \frac{f}{r^2} \right)''
-40fV'\left( \frac{f}{r^2} \right)'   \right. \nn && \left.
+3f'^{2}\left( \frac{f}{r^2} \right)''
+12f'f\left( \frac{f}{r^2} \right)'''
-40f\left( \frac{f}{r^2} \right)'\left( \mu^2 \frac{f}{r^2} \right)'
+2ff'''\left( \frac{f}{r^2} \right)'
-20Vf'\left( \frac{f}{r^2} \right)'    \right. \nn && \left.
+4\left (f\right )^{2}\left( \frac{f}{r^2} \right)''''
-40Vf\left( \frac{f}{r^2} \right)''\right] {\bf S^1_3(\varepsilon,\mu)}
-\frac{r^4}{128 f^3}
\left[4 f^{2}\left( \mu^2 \frac{f}{r^2} \right)''''
+f'f''\left( \mu^2 \frac{f}{r^2} \right)' \right. \nonumber
\ear
\bear
&&  \left.
-40fV'\left( \mu^2 \frac{f}{r^2} \right)'
-20Vf'\left( \mu^2 \frac{f}{r^2} \right)'
+8ff''\left( \mu^2 \frac{f}{r^2} \right)''
-20f\left( \mu^2 \frac{f}{r^2} \right)'^{2} \right. \nn && \left.
-40Vf\left( \mu^2 \frac{f}{r^2} \right)''
+3f'^{2}\left( \mu^2 \frac{f}{r^2} \right)''
+12f'f\left( \mu^2 \frac{f}{r^2} \right)'''
+2ff'''\left( \mu^2 \frac{f}{r^2} \right)'\right] {\bf S^0_3(\varepsilon,\mu)} \nonumber
\ear
\bear
&&
-\frac{r^6}{512 f^4}
\left[
-84 f^{2}\left( \frac{f}{r^2} \right)''^{2}
-112 f^{2}\left( \frac{f}{r^2} \right)'''\left( \frac{f}{r^2} \right)'
-21f'^{2}\left( \frac{f}{r^2} \right)'^{2}
-56ff''\left( \frac{f}{r^2} \right)'^{2}  \right. \nn && \left.
+280Vf\left( \frac{f}{r^2} \right)'^{2}
-252f'f\left( \frac{f}{r^2} \right)'\left( \frac{f}{r^2} \right)''\right] {\bf S^2_4(\varepsilon,\mu)}
-\frac{r^6}{256 f^4}
\left[
-56ff''\left( \frac{f}{r^2} \right)'\left( \mu^2 \frac{f}{r^2} \right)'
\right. \nonumber
\ear
\bear
&&
-21f'^{2}\left( \frac{f}{r^2} \right)'\left( \mu^2 \frac{f}{r^2} \right)'
+280Vf\left( \frac{f}{r^2} \right)'\left( \mu^2 \frac{f}{r^2} \right)'
-56 f^{2}\left( \mu^2 \frac{f}{r^2} \right)'''\left( \frac{f}{r^2} \right)'  \nn &&
-126f'f\left( \mu^2 \frac{f}{r^2} \right)'\left( \frac{f}{r^2} \right)''
-84\left (f\right )^{2}\left( \frac{f}{r^2} \right)''\left( \mu^2 \frac{f}{r^2} \right)''
-126f'f\left( \frac{f}{r^2} \right)'\left( \mu^2 \frac{f}{r^2} \right)''    \nn && \left.
-56\left (f\right )^{2}\left( \frac{f}{r^2} \right)'''\left( \mu^2 \frac{f}{r^2} \right)'\right] {\bf S^1_4(\varepsilon,\mu)}
-\frac{r^6}{512 f^4}
\left[280Vf\left( \mu^2 \frac{f}{r^2} \right)'^{2}\left (\frac{f}{r^2}\right )^{2} \right. \nn &&
-112f ^{2}\left( \mu^2 \frac{f}{r^2} \right)'\left( \mu^2 \frac{f}{r^2} \right)'''\left (\frac{f}{r^2}\right )^{2}
-252f'f\left( \mu^2 \frac{f}{r^2} \right)'\left( \mu^2 \frac{f}{r^2} \right)''\left (\frac{f}{r^2}\right )^{2}  \nonumber
\ear
\bear
&&  \left.
-21f'^{2}\left( \mu^2 \frac{f}{r^2} \right)'^{2}\left (\frac{f}{r^2}\right )^{2}
-56ff''\left( \mu^2 \frac{f}{r^2} \right)'^{2}\left (\frac{f}{r^2}\right )^{2}
-84 f^{2}\left( \mu^2 \frac{f}{r^2} \right)''^{2}\left (\frac{f}{r^2}\right )^{2}\right] {\bf S^0_4(\varepsilon,\mu)} \nn &&
-\frac{231\,r^8}{512f^4}\left[2f\left( \frac{f}{r^2} \right)'^{2}\left( \frac{f}{r^2} \right)''
+f\left( \frac{f}{r^2} \right)'^{3}f'\right] {\bf S^3_5(\varepsilon,\mu)}
-\frac{231\,r^8}{512f^4}
\left[3f'\left( \frac{f}{r^2} \right)'^{2}\left( \mu^2 \frac{f}{r^2} \right)'   \right. \nonumber
\ear
\bear
&&  \left.
+4f\left( \frac{f}{r^2} \right)''\left( \frac{f}{r^2} \right)'\left( \mu^2 \frac{f}{r^2} \right)'
+2f\left( \mu^2 \frac{f}{r^2} \right)''\left( \frac{f}{r^2} \right)'^{2}\right] {\bf S^2_5(\varepsilon,\mu)} \nn &&
-\frac{231\,r^8}{512f^4}
\left[2f\left( \frac{f}{r^2} \right)''\left( \mu^2 \frac{f}{r^2} \right)'^{2}
+4f\left( \mu^2 \frac{f}{r^2} \right)''\left( \frac{f}{r^2} \right)'\left( \mu^2 \frac{f}{r^2} \right)' \right. \nn && \left.
+3f'\left( \frac{f}{r^2} \right)'\left( \mu^2 \frac{f}{r^2} \right)'^{2}\right] {\bf S^1_5(\varepsilon,\mu)}
-\frac{231\,r^8}{512f^4}
\left[2f\left( \mu^2 \frac{f}{r^2} \right)''\left( \mu^2 \frac{f}{r^2} \right)'^{2}   \right. \nn && \left.
+f'\left( \mu^2 \frac{f}{r^2} \right)'^{3}\right] {\bf S^0_5(\varepsilon,\mu)}
+{\frac {1155\, r^{10}}{2048 f^4}}\left( \frac{f}{r^2} \right)'^{4} {\bf S^4_6(\varepsilon,\mu)}  \nn &&
+{\frac {1155\, r^{10}}{512 f^4}}\left( \frac{f}{r^2} \right)'^{3}\left( \mu^2 \frac{f}{r^2} \right)' {\bf S^3_6(\varepsilon,\mu)}
+{\frac {3465\, r^{10}}{1024 f^4}}\left( \frac{f}{r^2} \right)'^{2}\left( \mu^2 \frac{f}{r^2} \right)'^{2} {\bf S^2_6(\varepsilon,\mu)} \nn && \left.
+{\frac {1155\, r^{10}}{512 f^4}}\left( \frac{f}{r^2} \right)'\left( \mu^2 \frac{f}{r^2} \right)'^{3} {\bf S^1_6(\varepsilon,\mu)}
+{\frac {1155\,r^{10}}{2048 f^4}}\left( \mu^2 \frac{f}{r^2} \right)'^{4} {\bf S^0_6(\varepsilon,\mu)}
\right\}
\ear

\noindent
The mode sums and integrals in these expressions are of the form

      \bear \label{intsums}
      && S^m_k(\varepsilon,\mu) = \frac{\pi T r}{\sqrt{f}}
      \sum \limits_{l=0}^{\infty}  \left\{  \frac{\left(l+1/2\right)^{2m+1}}
      {\left[\mu^2+\left(l+1/2\right)^2\right]^{(2k+1)/2}}-
      \mbox{subtraction terms} \right\} \nn &&
      +\frac{2 \pi T r}{\sqrt{f}}\sum \limits_{n=1}^{\infty} \cos (\varepsilon u_n)
      \sum \limits_{l=0}^{\infty}  \left\{  \frac{\left(l+1/2\right)^{2m+1}}
      {\left[u_n^2+\mu^2+\left(l+1/2\right)^2\right]^{(2k+1)/2}}-
      \mbox{subtraction terms} \right\},
      \ear

\noindent
where $m$ and $k$ are integers  and $m, k \geq 0$.
The subtraction terms for the sum over $l$  can be obtained by expanding
the function that is summed in inverse powers of $l$ and truncating at
$O(l^0)$. Such subtracting corresponds to removing the superficial
divergences in the sums over $l$ discussed above:
      \bear \label{Smm}
      S^m_m(\varepsilon,\mu) &=& \frac{\pi T r}{\sqrt{f}}
      \sum \limits_{l=0}^{\infty}  \left\{  \frac{\left(l+1/2\right)^{2m+1}}
      {\left[\mu^2+\left(l+1/2\right)^2\right]^{(2m+1)/2}}-
      1 \right\} \nn &&
      +\frac{2 \pi T r}{\sqrt{f}}\sum \limits_{n=1}^{\infty} \cos (\varepsilon u_n)
      \sum \limits_{l=0}^{\infty}  \left\{  \frac{\left(l+1/2\right)^{2m+1}}
      {\left[u_n^2+\mu^2+\left(l+1/2\right)^2\right]^{(2m+1)/2}}-
      1 \right\}.
      \ear

\noindent
For other quantities $S^m_k(\varepsilon, \mu)$ there are no subtraction terms.
The details of calculations of $S^m_k(\varepsilon,\mu)$ in the limit
$\varepsilon \rightarrow 0$  are discussed in Appendix A:
     \bear \label{start}
     S^{0}_{0}(\varepsilon, \mu)&=&\frac{1}{\varepsilon^2}
     +4 \mu^2 J\left({a}/{\mu}\right)
     +\left( \frac{\mu^2}{2}-\frac{1}{24} \right)
     \left( C+\frac12\ln\left| \frac{\mu^2 \varepsilon^2}{4} \right|\right)
     \nn &&
     -\frac{\mu^2}{4}-\mu^2 I_1(\mu)
     +O\left(\varepsilon^2 \ln \left|\varepsilon \right| \right),
     \ear
     \bear
     S^{0}_{1}(\varepsilon, \mu)&=&-\left( C+\frac12\ln\left|
     \frac{\mu^2 \varepsilon^2}{4} \right|\right)+
     \mu \frac{d}{d \mu}I_0(\mu)
     +O\left(\varepsilon^2 \ln \left|\varepsilon \right| \right),
     \ear
     \bear
     S^{m}_{m+1}(\varepsilon, \mu)&=&\frac{1}{(2m+1)!!}\left[-(2 m)!!
     \left( C+\frac12\ln\left|\frac{\mu^2 \varepsilon^2}{4} \right|\right)
     +\mu^2\left( \frac{\partial}{\mu \partial\mu} \right)^{m+1}
     \left(\mu^{2m}I_0(\mu)\right) \right] \nn
     &&+O\left(\varepsilon^2 \ln \left|\varepsilon \right| \right) \quad (m\geq 1),
     \ear
     \bear\label{finish}
     S^{m}_{k}(\varepsilon, \mu)&=&\frac{(2m+1)!!}{(2k+1)!!}\left(
     -\frac{\partial}{\mu \partial \mu} \right)^{k-m-1} S^{m}_{m+1}
     (\varepsilon, \mu) \quad (m\geq 0, \ k \geq m+2),
     \ear
where
     \beq
     J\left({a}/{\mu}\right)=\int^\infty_1 \frac{\sqrt{\eta^2-1}}{e^{2 \pi \mu \eta / a}-1} d \eta, \quad
     I_m(\mu)=\int \nolimits_{0}^{\infty}\frac{\eta^{2m-1}\ln|1-\eta^2|}
     {1+e^{2\pi \mu \eta}}d\eta.
     \eeq

The renormalization of $\langle \varphi^2 \rangle$ is achieved by subtracting the
renormalization counterterm from $\langle \varphi^2 \rangle_{unren}$ and
then letting $\tilde \tau \rightarrow \tau$:
      \beq
      \left< \varphi^2 \right>_{ren}=
      \lim_{\tilde \tau \rightarrow \tau}
      \left[\left< \varphi^2 \right>_{unren}-
      \left< \varphi^2 \right>_{\mbox{\tiny DS}}\right],
      \eeq
where
      \bear
      \langle \phi^2 \rangle_{\mbox{\tiny DS}} &=& \frac1{8\pi^2\sigma}+\frac1{8\pi^2}
      \left[m^2+\left(\xi-\frac16\right)R\right]
      \left[C+\frac12\ln\left( \frac{m_{ \mbox{\tiny \sl DS}}^2|\sigma|}{2}
      \right)\right] \nn && -\frac{m^2}{16\pi^2}
      +\frac1{96\pi^2}R_{\alpha\beta}
      \frac{\sigma^\alpha\sigma^\beta}{\sigma},
      \ear

\noindent
$C$ is Euler's constant, $R_{\alpha\beta}$ is the Ricci tensor,
$R=R^{\alpha}_{\alpha}$, $\sigma$ is one-half the square of the
distance between the points $x$ and $\tilde x$ along the shortest
geodesic connecting them, $\sigma^{\mu}$ is the covariant
derivative of $\sigma$ \cite{Popov:2007}.
\begin{eqnarray}
        &&{\sigma^i}= -\left(x^i-\tilde x^i\right)
        -\frac12 \Gamma^{i}_{{j}{k}}\left(x^j-{\tilde x^j}\right)\left(x^k-{\tilde x^k}\right)
        \nonumber \\ &&
        -\frac16 \left( \Gamma^{i}_{{j}{m}} \Gamma^{m}_{{k}{l}}
        +\frac{\partial \Gamma^{i}_{{j}{k}}}{\partial {\tilde x^l}}\right)
        \left(x^j-{\tilde x^j}\right)\left(x^k-{\tilde x^k}\right)\left(x^l-{\tilde x^l}\right)
        %\nonumber \\ &&
        +O\left(\left(x-{\tilde x}\right)^4\right),\nn
      &&\sigma = \frac12 g_{\mu \nu}\sigma^{\mu}\sigma^{\nu},
        \end{eqnarray}
where Christoffel symbols $\Gamma^{i}_{{j}{k}}$ are calculated at point $\tilde x$.
For the metric (\ref{metric}) the quantities $\sigma^{\mu}$ and $\sigma$
are
      \bear
      \sigma^\tau&=&-(\tau-\tilde \tau)+\frac{f'^2}{24f}(\tau-\tilde \tau)^3
      +O\left((\tau-\tilde \tau)^5\right),\nn
      \sigma^{\rho}&=&\frac{f'}{4}(\tau-\tilde \tau)^2
      +O\left((\tau-\tilde \tau)^4\right),\nn
      \sigma^{\theta}&=&\sigma^{\phi}=0.
      \ear
The constant $m_{ \mbox{\tiny \sl DS}}$ is equal to the
mass $m$ of the field for a massive scalar field. For a massless
field $m_{ \mbox{\tiny \sl DS}}$ is an arbitrary parameter due to the infrared cutoff
in $\left< \varphi^2\right>_{\mbox{\tiny DS}}$. A particular
choice of the value of $m_{ \mbox{\tiny \sl DS}}$ corresponds to a finite
renormalization of the coefficients of terms in the gravitational
Lagrangian and must be fixed by experiment or observation.
\bear
\langle \phi^2 \rangle_{\mbox{\tiny DS}} &=& \frac1{8\pi^2}\left\{\frac{2}{f (\tau-\tilde \tau)^2} +\frac{{f'}^2}{f^2}
+\left[m^2+\left(\xi-\frac16\right)\left( \frac{2}{r^2}-\frac{f''}{f}-\frac{2 {r^2}''}{r^2} +\frac{{f'}^2}{2 f^2} -\frac{f' {r^2}'}{f r^2} \right. \right. \right.
\nn
&&\left. \left. \left. +\frac{{{r^2}'}^2}{2r^4} \right)\right] \left[C+\frac12\ln\left( \frac{m_{ \mbox{\tiny \sl DS}}^2 f (\tau-\tilde \tau)^2}{4} \right)\right]
-\frac{m^2}{2} -\frac{f' {r^2}'}{12 f r^2} -\frac{f''}{12 f} \right\}.
\ear
The procedure described above gives renormalized expression for
$\left< \varphi^2 \right>_{ren}$ in the framework of the second-order
WKB approximation:
\bear
&&4 \pi^2 \left< \varphi^2 \right>_{ren}  = 4 \pi^2 \left(
\left< \varphi^2 \right>^{(0)}_{ren} +\left< \varphi^2 \right>^{(2)}_{ren}  \right) \nn &&
= \frac{4 \mu^2}{r^2} J\left({a}/{\mu}\right)
-\frac{1}{4 r^2} \left( 2 \xi -\frac{1}{4} \right)
+\frac{1}{4 r^2} \left[ m^2 r^2 +2\left(\xi-\frac{1}{6} \right) \right] \ \ln \left| \frac{m^2 r^2 +2 \xi -1/4}{m_{ \mbox{\tiny \sl DS}}^2 r^2} \right| \nn
&& -\frac{\mu^2}{r^2} I_1(\mu)+\frac{{r^2}''}{24 r^2}-\frac{{{r^2}'}^2}{48 r^4}+\frac{f' {r^2}'}{48 f r^2}
+ \left[ \frac{f''}{12 f} +\frac{{r^2}''}{6 r^2} -\frac{{f'}^2}{24 f^2} -\frac{{{r^2}'}^2}{24 r^4} +\frac{f' {r^2}'}{12 f r^2}
\right.  \nn &&  \left.
+\xi\left( -\frac{f''}{2 f} -\frac{{r^2}''}{r^2} +\frac{{f'}^2}{4 f^2}
+\frac{{{r^2}'}^2}{4 r^4} -\frac{f' {r^2}'}{2 f r^2} \right)
\right] \left(\frac12 \ln \left| \frac{m^2 r^2 +2 \xi -1/4}{m_{ \mbox{\tiny \sl DS}}^2 r^2} \right| -
\mu \frac{d I_0(\mu)}{d \mu} \right),
\ear
where $a={2 \pi T r}/{\sqrt{f}}$.
The terms of a fourth WKB order are given in Appendix B.

%--------------------------------------------------------------------

\section{Conclusion}

%--------------------------------------------------------------------

We have obtained an analytical approximation for $\langle \varphi^2 \rangle_{ren}$ of quantized scalar fields
in the background of a long throat (\ref{lwkb})-(\ref{Lm}).
The field is assumed to be both massive or massless, with arbitrary coupling to the scalar curvature, and in
a thermal state at an arbitrary temperature.

As an example, we consider spacetime with metric
\begin{equation} \label{UEH1}
ds^{2}=-f(r)dt^{2}+\frac{dr^{2}}{f(r)}+r^{2}\left( d\theta ^{2}+\sin^{2}\theta d\varphi ^{2}\right),
\end{equation}
where $f(r)$ is
\begin{equation} \label{UEH2}
f(r)= -\frac{\left( r+3r_{+}\right) }{6{r}_{+}\!{}^{2}\,r^{2}}\left( r-r_{+}\right) ^{3}.
\end{equation}
It was shown in \cite{PopZas:2007} the WKB approximation (\ref{lwkb})-(\ref{Lm}) remains valid near
ultraextremal horizon $r=r_+$
It means that we can obtain explicit  expression for $\langle \varphi^2 \rangle_{ren}$ near
ultraextremal horizon $r=r_+$
\bear
&&4 \pi^2 \left< \varphi^2 \right>_{ren} = \frac{4 \mu^2}{r_+^2} J\left({a}/{\mu_+}\right)-\frac{1}{4 r_+^2} \left( 2 \xi -\frac{1}{4} \right)
+\frac{1}{4 r_+^2} \left[ m^2 r_+^2
\right. \nn && \left.
+2\left(\xi-\frac{1}{6} \right) \right] \ \ln \left| \frac{\mu_+^2}{m_{ \mbox{\tiny \sl DS}}^2 r^2} \right|
-\frac{\mu_+^2}{r_+^2} I_1(\mu_+)
+\left[
\frac{2\xi-1/4}{24 r_+^3 \mu_+^2} +2 \ \frac{2\xi-1/4}{r_+^3} I_1(\mu_+)
\right. \nn && \left.
+\frac{m^2 \mu_+}{r_+} \ \frac{d I_1(\mu_+)}{d \mu_+}
+\frac{2 \mu_+}{r_+^3} \left( \xi-\frac16 \right) \frac{d I_0(\mu_+)}{d \mu_+}
\right] \left( r - r_+ \right) \nn
&& +\left[
\frac{-128 m^4 r_+^4 +52 m^2 r_+^2 -3 -480 m^2 r_+^2 \xi +64 \xi -320 \xi^2}{192 r_+^4 \mu_+^4}
\right. \nn
&&
-3 \frac{2 \xi-1/4}{r_+^4} I_1(\mu_+)
 +\frac{3 m^2 (2 \xi-1/4)}{2 \mu_+ r_+^2} \frac{d I_1(\mu_+)}{d \mu_+}
-\frac{m^4}{2} \frac{d^2 I_1(\mu_+)}{d \mu_+^2}
\nn && \left.
+ \frac{4 m^2 r_+^2 -3 -24 m^2 r_+^2 \xi +50 \xi -192 \xi^2}{24 r_+^4 \mu_+} \ \frac{d I_0(\mu_+)}{d \mu_+} \right. \nn
&& \left. + \frac{- m^2 r_+^2  +6 m^2 r_+^2 \xi + \xi -6 \xi^2}{3 r_+^4 } \ \frac{d^2 I_0(\mu_+)}{d \mu_+^2}
\right] \left( r - r_+ \right)^2 +O\left( \left( r - r_+ \right)^3 \right),
\ear
where $a_+=\frac{\dst 2 \pi T r_+}{\dst \sqrt{f(r_+)}}, \mu_+^2 = {m^2 r_+^2 +2 \xi -1/4}$.
%--------------------------------------------------------------------
\acknowledgments{The work is performed according to the Russian Government Program of Competitive Growth of Kazan Federal University.}

%--------------------------------------------------------------------
\section*{Appendix A}

\setcounter{equation}{0}
\renewcommand{\theequation}{A\arabic{equation}}
%--------------------------------------------------------------------
To calculate the quantities $S^m_k(\varepsilon, \mu)$ it is
necessary to compute the various sums over $l$. We start from the
sums in expression (\ref{intsums}):
      \bear \label{Smn}
      && S^m_k(\varepsilon,\mu) = \frac{\pi T r}{\sqrt{f}}
      \sum \limits_{l=0}^{\infty}  \left\{  \frac{\left(l+1/2\right)^{2m+1}}
      {\left[\mu^2+\left(l+1/2\right)^2\right]^{(2k+1)/2}}-
      \mbox{subtraction terms} \right\} \nn &&
      +\frac{2 \pi T r}{\sqrt{f}}\sum \limits_{n=1}^{\infty} \cos (\varepsilon u_n)
      \sum \limits_{l=0}^{\infty}  \left\{  \frac{\left(l+1/2\right)^{2m+1}}
      {\left[u_n^2+\mu^2+\left(l+1/2\right)^2\right]^{(2k+1)/2}}-
      \mbox{subtraction terms} \right\},
      \ear
For calculation of this sum we can use the Abel-Plana method
\cite{Evg,Sushkov2,P}
      \beq
      \sum \limits_{l=0}^{\infty} F( l+1/2)
      =\int \nolimits_{q}^{\infty}F(x)dx
      +\int \nolimits_{q-i\infty}^{q}\frac{F(z)}{1+e^{i2\pi z}}dz
      -\int \nolimits_{q}^{q+i\infty}\frac{F(z)}{1+e^{-i2\pi z}}dz,
      \eeq
where $-1/2<q<1/2$, $f(z)$ is a holomorphic function for $Re z\geq
q$, $f(z)$ satisfies the condition
      \beq \label{1a}
      \left| F(x+iy) \right|<\epsilon(x)e^{a|y|}, \quad 0<a<2\pi,
      \eeq
and $\epsilon(x)$ is an arbitrary function with asymptotic
behavior
      \beq
      \epsilon(x)\rightarrow 0 \quad \mbox{for} \quad x
      \rightarrow +\infty.
      \eeq
Using this formula we can calculate the sums in Eqs. (\ref{intsums}):
      \bear
      &&\sum \limits_{l=0}^{\infty}  \left\{  \frac{\left(l+1/2\right)^{2m+1}}
      {\left[u_n^2+\mu^2+\left(l+1/2\right)^2\right]^{(2m+1)/2}}-1\right\}
     =\lim _{q\rightarrow 0}\left\{ \int \nolimits_{q}^{\infty} \left[
     \frac{x^{2m+1}}{\left(u_n^2+\mu^2+x^2\right)^{(2m+1)/2}}-1\right]dx\right.
     \nn && +\int \nolimits_{q-i\infty}^{q} \left[ \frac{z^{2m+1}}
    {\left(u_n^2+\mu^2+z^2\right)^{(2m+1)/2}}-1\right]\frac{dz}{(1+e^{i2\pi z})}
     -\int \nolimits_{q}^{q+i\infty} \left[ \frac{z^{2m+1}}
     {\left(u_n^2+\mu^2+z^2\right)^{(2m+1)/2}} \right. \nn && \left. \left.
     -1 \frac{}{} \right]\frac{dz}{(1+e^{-i2\pi z})}
      \right\}=\sum \nolimits_{k=0}^{m}\frac{(-1)^{m+k}m!}{k!(m-k)!}
      \frac{\sqrt{u_n^2+\mu^2}}{(2m-2k-1)}\nn
      &&+2(-1)^m\lim _{\delta\rightarrow 0}\left[\int \nolimits_{0}^{
      \sqrt{u_n^2+\mu^2}-\delta}\frac{x^{2m+1}}{(1+e^{2\pi x})
      \left(u_n^2+\mu^2-x^2\right)^{(2m+1)/2}}dx \right. \nn && \left.
      -\left(\mbox{terms of this integral that}\atop\mbox
      { diverge in the limit}\ \delta \rightarrow 0  \right) \right]
      =\sum \nolimits_{i=0}^{m}\frac{(-1)^{m+i}m!}{i!(m-i)!}
      \frac{\sqrt{u_n^2+\mu^2}}{(2m-2i-1)}
      \nn && +\frac{2}{(2m-1)!!}
      \int \nolimits_{0}^{\sqrt{u_n^2+\mu^2}}\frac{xdx}{\sqrt{u_n^2+\mu^2-x^2}}
      \left( \frac{d}{xdx} \right)^m \frac{x^{2m}}{(1+e^{2\pi x})}
      \quad (m \geq 0),
      \ear
      \bear
      &&\sum \limits_{l=0}^{\infty} \frac{\left(l+1/2\right)^{2m+1}}
      {\left[u_n^2+\mu^2+\left(l+1/2\right)^2\right]^{(2m+3)/2}}
      =\lim _{q\rightarrow 0}\left\{
      \int\nolimits_{q}^{\infty}\frac{x^{2m+1}}{\left(u_n^2+\mu^2+x^2
      \right)^{(2m+3)/2}}dx\right.\nn
      &&\left.+\int \nolimits_{q-i\infty}^{q}  \frac{z^{2m+1}}
      {(u_n^2+\mu^2+z^2)^{(2m+3)/2}}\frac{dz}
      {(1+e^{i2\pi z})}-\int \nolimits_{q}^{q+i\infty}\frac{z^{2m+1}}
      {(u_n^2+\mu^2+z^2)^{(2m+3)/2}}\frac{dz}{(1+e^{-i2\pi z})}\right\}\nn
      &&=\frac{2^m m!}{(2m+1)!!\sqrt{u_n^2+\mu^2}}
      +2(-1)^m\lim _{\delta\rightarrow 0}\left[\int
      \nolimits_{0}^{\sqrt{u_n^2+\mu^2}-\delta}\frac{x^{2m+1}}
      {(u_n^2+\mu^2-x^2)^{(2m+3)/2}}\frac{dx}{(1+e^{2\pi x})}\right.\nn
      &&\left.-\left(\mbox{terms of this integral that}\atop\mbox
      { diverge in the limit}\ \delta \rightarrow 0  \right) \right] \nn &&
      =-\frac{2}{(2m+1)!!}\int \nolimits_{0}^{\sqrt{u_n^2+\mu^2}}
      \frac{xdx}{\sqrt{u_n^2+\mu^2-x^2}}\left( \frac{d}{xdx} \right)^{m+1}
      \frac{x^{2m}}{1+e^{2\pi x}}
      \quad (m\geq 0),
      \ear
      \bear
      &&\sum \limits_{l=0}^{\infty} \frac{\left(l+1/2\right)^{2m+1}}
      {\left[u_n^2+\mu^2+\left(l+1/2\right)^2\right]^{(2k+1)/2}} \nn &&
      =\frac{(2m+1)!!}{(2k-1)!!}\left( -\frac{\partial }{\mu\partial \mu}
   \right)^{k-m-1}\sum \limits_{l=0}^{\infty} \frac{\left(l+1/2\right)^{2m+1}}
      {\left[u_n^2+\mu^2+\left(l+1/2\right)^2\right]^{(2m+3)/2}}\nn
     &&=\frac{2 (-1)^{k-m}}{(2k-1)!!}\left( \frac{\partial }{\mu\partial \mu}
      \right)\int \nolimits_{0}^{\sqrt{u_n^2+\mu^2}}
      \frac{xdx}{\sqrt{u_n^2+\mu^2-x^2}}\left( \frac{d}{xdx} \right)^{m+1}
      \frac{x^{2m}}{1+e^{2\pi x}} \quad
      \left(m \geq 0, \atop k \geq m+2\right).
      \ear

Now we can calculate the sums over $n$ in (\ref{intsums}).
      \bear \label{An}
      \sum_{n=1}^{\infty} \sqrt{u_n^2+\mu^2} \cos (\varepsilon u_n)
      &=&\left( -\frac{d^2}{d\varepsilon^2}+\mu^2 \right)\sum_{n=1}^{\infty} \frac{\cos (\varepsilon a n) }{\sqrt{a^2 n^2+\mu^2}}
      \nn
       &=& a\left( -\frac{d^2}{d\tilde \varepsilon^2}+{\tilde \mu}^2 \right)\left[ -\frac{1}{\tilde \mu}+\sum_{n=0}^{\infty} \frac{\cos (\tilde \varepsilon  n) }{\sqrt{n^2+{\tilde \mu}^2}} \right],
      \ear
where
      \beq
      a=\frac{2 \pi T r}{\sqrt{f}}, \ \tilde \varepsilon=a \varepsilon, \ \tilde \mu= \frac{\mu}{a}.
      \eeq
For calculation of the last sum over $n$ we can use the Plana sum formula \cite{21}
      \beq \label{AP}
      \sum _{n=0}^{\infty} F(n)=
      \frac{1}{2} F(0) + \lim _{q\rightarrow 0}\left\{ \int_{q}^{\infty} F(y) dy +i \int_{q}^{\infty} \frac{d y}{e^{2 \pi y}-1} \left[\frac{}{} F(q+i y) - F(q-i y)  \right] \right\}.
      \eeq
Consequently
      \bear
      \sum_{n=0}^{\infty} \frac{\cos (\tilde \varepsilon  n) }{\sqrt{n^2+{\tilde \mu}^2}}
      =\frac{1}{ 2 \tilde \mu} + \int_0^\infty \frac{\cos (\tilde \varepsilon y) }{\sqrt{y^2+{\tilde \mu}^2}} d y
      +2 \int_{\tilde \mu}^\infty \frac{\cosh (\tilde \varepsilon y) }{\sqrt{y^2-{\tilde \mu}^2}} \frac{d y}{(e^{2 \pi y}-1)}.
      \ear
Note that the sum (\ref{An}) can be expanded in powers of $\varepsilon$:
      \bear
      &&\sum_{n=1}^{\infty} \sqrt{u_n^2+\mu^2} \cos (\varepsilon u_n)
      = a\left( -\frac{d^2}{d\tilde \varepsilon^2}+{\tilde \mu}^2 \right) \left[-\frac{1}{2 \tilde \mu} +K_0(\tilde \varepsilon \tilde \mu)
      \right. \nn && \left.
      +2 \int_{\tilde \mu}^\infty \frac{\cosh (\tilde \varepsilon y) }{\sqrt{y^2-{\tilde \mu}^2}} \frac{d y}{(e^{2 \pi y}-1)} \right]
      = a\left[-\frac{\tilde \mu}{2}-\frac{\tilde \mu}{\tilde \varepsilon} K_1(\tilde \varepsilon \tilde \mu)
      \right. \nn && \left.
      -2 \int_{\tilde \mu}^\infty \frac{\cosh (\tilde \varepsilon y) }{\sqrt{y^2-{\tilde \mu}^2}} \frac{y^2 d y}{(e^{2 \pi y}-1)}
      +2 {\tilde \mu}^2 \int_{\tilde \mu}^\infty \frac{\cosh (\tilde \varepsilon y) }{\sqrt{y^2-{\tilde \mu}^2}} \frac{d y}{(e^{2 \pi y}-1)}
      \right]
       \nn &&
      = a\left[-\frac{\tilde \mu}{2}
      -\frac{{\tilde \mu}^2}{2}
      \left(C+\frac12\ln\frac{(\tilde \varepsilon \tilde \mu)^2}{4}\right)
       -\frac{1}{{\tilde \varepsilon}^2}+\frac{{\tilde \mu}^2}{4}+O\left({\tilde \varepsilon}^2 \ln \left|\tilde \varepsilon
      \right| \right) \right]
      \nn &&
      -2 a \int_{\tilde \mu}^\infty \frac{\sqrt{y^2-{\tilde \mu}^2} d y}{e^{2 \pi y}-1},
      \ear
where $K_n(x)$ is Macdonald's function and $C$ is Euler's constant.

The second type of sum over $n$ has the form
      \bear
      I_{-} &=&\sum_{n=1}^{\infty} \cos (\varepsilon u_n)
      \int \nolimits_{0}^{\sqrt{u_n^2+\mu^2}}
      \frac{G(x)}{\sqrt{ u_n^2+\mu^2-x^2}} \ dx.
      \ear
Using the formula (\ref{AP}) one can obtain
      \bear
      &&I_{-}=
      -\frac12 \int \nolimits_{0}^{\tilde \mu}\frac{G(\tilde x a)}{\sqrt{ \tilde \mu^2 -\tilde x^2}} \ d \tilde x
      + \lim _{q\rightarrow 0} \left\{ \int \nolimits_{q}^{\infty} dy \cos (\tilde \varepsilon y)
      \int \nolimits_{0}^{\sqrt{y^2+\tilde \mu^2}}
      \frac{G(\tilde x a)}{\sqrt{ y^2+\tilde \mu^2 -\tilde x^2}} \ d \tilde x
      \right. \nn  &&
      +i \int \nolimits_{q}^{\infty} \frac{dy \cos [\tilde \varepsilon (q+iy)]}{e^{2 \pi y}-1}
      \int \nolimits_{0}^{\sqrt{(q+iy)+\tilde \mu^2}} \frac{G(\tilde x a) d \tilde x}{\sqrt{(q+iy)^2+ \tilde \mu^2 -\tilde x^2}}
      \nn && \left.
      -i \int \nolimits_{q}^{\infty} \frac{dy \cos [\tilde \varepsilon (q-iy)]}{e^{2 \pi y}-1}
      \int_{0}^{\sqrt{(q-iy)+\tilde \mu^2}} \frac{G(\tilde x a) d \tilde x}{\sqrt{(q-iy)^2 +\tilde \mu^2 -\tilde x^2}} \right\}.
      \ear
Changing the order of integration over $y$ and $\tilde x = x/a$ gives
      \bear
      &&\int \nolimits_{0}^{\infty} dy \cos (\tilde \varepsilon y)
      \int \nolimits_{0}^{\sqrt{y^2+\tilde \mu^2}}
      \frac{G(\tilde x a)}{\sqrt{ y^2+\tilde \mu^2 -\tilde x^2}} \ d \tilde x =
      \int \nolimits_{0}^{\tilde \mu} d \tilde x \, G(\tilde x a)\int \nolimits_{0}^{\infty}dy
      \frac{\cos (\tilde \varepsilon \tilde y)}{\sqrt{y^2+\tilde \mu^2 -\tilde x^2}}
      \nn &&
      +\int \limits_{\tilde \mu}^{\infty}d\tilde x \, G(\tilde x a) \int \limits_{\sqrt{{\tilde x}^2 -\tilde \mu^2}}^{\infty}
      \frac{\cos (\tilde \varepsilon y)}{\sqrt{y^2 +\tilde \mu^2 -\tilde x^2}} dy.
      \ear
Since
      \bear \label{A}
      &&\int \nolimits_{0}^{\infty}\frac{\cos (\tilde \varepsilon y) dy}
      {\sqrt{y^2+\tilde \mu^2 -\tilde x^2}}=K_0(\tilde \varepsilon \sqrt{\tilde \mu^2 -\tilde x^2})\nn
      && = -\left(C+\frac12\ln\left|\frac{\tilde \varepsilon^2 (\tilde \mu^2 -\tilde x^2)}{4}
      \right|\right)
      +O\left(\tilde \varepsilon^2 \ln \left|\tilde \varepsilon \right| \right),
      \ear
      \bear
      &&\int \nolimits_{\sqrt{\tilde x^2 -\tilde \mu^2}}^{\infty}\frac{\cos (\tilde \varepsilon y) dy}
     {\sqrt{y^2+\tilde \mu^2 -\tilde x^2}}=-\frac{\pi}{2}N_0(\tilde \varepsilon \sqrt{\tilde x^2 -\tilde \mu^2})\nn
      && = - \left(C+\frac12\ln\left|\frac{\tilde \varepsilon^2(\tilde x^2- \tilde \mu^2)}{4}\right|\right)
      +O\left(\tilde \varepsilon^2 \ln \left| \tilde \varepsilon \right| \right),
      \ear
where  $N_0(x)$ is Neumann's function,  one can obtain
      \bear
      &&\int \nolimits_{0}^{\infty} dy \cos (\tilde \varepsilon y)
      \int \nolimits_{0}^{\sqrt{y^2+\tilde \mu^2}}
      \frac{G(\tilde x a)}{\sqrt{ y^2+\tilde \mu^2 -\tilde x^2}} \ d \tilde x \nn &&
      = -\int \nolimits_{0}^{\infty}d\tilde x G(\tilde x a)
      \left(C+\frac12\ln\left|\frac{\tilde \varepsilon^2(\tilde x^2- \tilde \mu^2)}{4} \right|\right)
      +O\left(\tilde \varepsilon^2 \ln \left|\tilde \varepsilon \right| \right)
       \nn &=&
      -\frac{1}{a}\int \nolimits_{0}^{\infty}d \eta \, G(x)
      \left(C+\frac12\ln\left|\frac{\varepsilon^2 \mu^2}{4}\right| +\frac12 \ln\left|\frac{x^2}{\mu^2}- 1 \right|\right)
      +O\left( \varepsilon^2 \ln \left| \varepsilon \right| \right).
      \ear
 Thus
      \bear
      I_- &=&-\frac12 \int \nolimits_{0}^{\mu}\frac{G(x)}{\sqrt{ \mu^2 - x^2}}\ d x
      -\frac{1}{a}\int \nolimits_{0}^{\infty}d x \, G(x) \left(C+\frac12\ln\left|\frac{\varepsilon^2 \mu^2}{4}\right| +\frac12 \ln\left|\frac{x^2}{\mu^2}- 1 \right|\right) \nn &&
      +\frac{1}{a}
      \int _{\mu}^{\infty} \frac{du}{e^{2 \pi u/a}-1}
      \int_{0}^{\sqrt{u^2-\mu^2}}  \frac{\left[ i G(iz) -i G(-iz) \right]}{\sqrt{u^2 - \mu^2 -z^2}} d z
      +O\left( \varepsilon^2 \ln \left| \varepsilon \right| \right).
      \ear
If we also take in consideration
      \beq
    \int \nolimits_{0}^{\infty}\frac{x dx}{1+e^{2\pi x}}=\frac{1}{48}, \quad
      \int \nolimits_{0}^{\infty}\frac{x^3 dx}{1+e^{2\pi x}}=\frac{7}{1920},
      \eeq
      \beq
      \frac{\partial}{\mu \partial \eta}G(\mu \eta) =\frac{\partial}
      {\eta \partial \mu}G(\mu \eta),
      \eeq
the resulting expressions for $S^{m}_{k}(\varepsilon, \mu)$ can be
presented as Eqs. (\ref{start})-(\ref{finish}).

%--------------------------------------------------------------------
\section*{Appendix B}

\setcounter{equation}{0}
\renewcommand{\theequation}{B\arabic{equation}}
%--------------------------------------------------------------------

The terms of a fourth WKB order of $\left< \varphi^2 \right>_{ren}$ are
\bear
&& 4 \pi^2 \left< \varphi^2 \right>^{(4)}_{ren} =
 \left\{  \left[
-{\frac {3 {{f'}^2} {{{{(r^2)}'}}^2}}{16 {f^2}{r^2}  }}
+{\frac {{r^2}   {{f'}^2}f''  }{8 {f^3}}}
-{\frac { {f'}  {{{(r^2)}'}} f''  }{4 {f^2}}}
-{\frac {{r^2}   {f'}^4}{32 {f^4}}}
-{\frac { {{(r^2)}''}  {f'} {{{(r^2)}'}}  }{2f  {r^2}  }}
\right. \right. \nn &&
-{\frac { {{{{(r^2)}'}}^4}}{32 {r^6}}}
+{\frac { {{{{(r^2)}'}}^2}f''  }{8f  {r^2}  }}
-{\frac {{r^2}   {f''} ^{2}}{8 {f^2}}}
-{\frac {{r^2}   {f''} ^{2}}{8 {f^2}}}
+{\frac { {{f'}^3}{{{(r^2)}'}}  }{8 {f^3}}}
+{\frac { {f'}  {{{{(r^2)}'}}^3}}{8f   {r^4}}}
+{\frac { {{(r^2)}''}  {{{{(r^2)}'}}^2}}{4 {r^4}}}
\nn && \left.
+{\frac { {{f'}^2}{{(r^2)}''}  }{4 {f^2}}}
-{\frac { {{(r^2)}''} ^{2}}{2{r^2}  }}
-{\frac { {{(r^2)}''} f''  }{2f  }} \right] {\xi}^{2}
+ \left[ {\frac {5{r^2}   {f'} f'''  }{48 {f^2}}}
+{\frac {3 {{(r^2)}''}  {f'} {{{(r^2)}'}}  }{16f  {r^2}  }}
+{\frac { {f'}  {{{(r^2)}'}} f''  }{3 {f^2}}}
\right. \nn &&
-{\frac {5 {f'}  {{{{(r^2)}'}}^3}}{96f   {r^4}}}
+{\frac { {{(r^2)}''} f''  }{24f  }}
+{\frac { {{{(r^2)}'}} {{(r^2)}'}''  }{24{r^2}  }}
-\frac{r^2 f''''}{24 f}
-{\frac {3 {{(r^2)}''}  {{{{(r^2)}'}}^2}}{16 {r^4}}}
+{\frac {5 {{{{(r^2)}'}}^4}}{96 {r^6}}}
-{\frac {17 {{f'}^3}{{{(r^2)}'}}  }{96 {f^3}}} \nonumber
\ear
\bear
&&
+{\frac {5 {{(r^2)}''} ^{2}}{24{r^2}  }}
-\frac{{(r^2)}''''}{12}
+{\frac {11{r^2}   {f'}^4}{96 {f^4}}}
-{\frac { {f'''} {{{(r^2)}'}}  }{8 f  }}
-{\frac { {f'} {{(r^2)}'}''  }{12 f  }}
+{\frac {{r^2}   {f''} ^{2}}{ {8 f^2}}}
+{\frac { {{f'}^2} {{{{(r^2)}'}}^2}}{16 {f^2}{r^2}  }} \nn
&& -{\frac {7{r^2}   {{f'}^2}f''  }{24 {f^3}}}
\left.-{\frac { {{{{(r^2)}'}}^2}f''  }{24f  {r^2}  }} \right] \xi
-{\frac {49 {f'}  {{{(r^2)}'}} f''  }{960 {f^2}}}
+{\frac { {{{{(r^2)}'}}^2}f''  }{480 f  {r^2}  }}
-{\frac { {{(r^2)}''}  {f'} {{{(r^2)}'}}  }{60 f  {r^2}  }}
+{\frac { {f'}  {{{{(r^2)}'}}^3}}{192 f   {r^4}}}  \nonumber
\ear
\bear
&&
-{\frac {3 {{(r^2)}''} ^{2}}{160 {r^2}  }}
-{\frac {11 {{{{(r^2)}'}}^4}}{1920 {r^6}}}
+{\frac {13 {{f'}^3}{{{(r^2)}'}}  }{480 {f^3}}}
-{\frac {3 {{f'}^2} {{{{(r^2)}'}}^2}}{640 {f^2}{r^2}  }}
+{\frac {11 {f'''} {{{(r^2)}'}}  }{480 f  }}
+{\frac { {f'} {{(r^2)}'}''  }{80 f  }}
-{\frac { {{{(r^2)}'}} {{(r^2)}'}''  }{240 {r^2}  }} \nn
&&
+{\frac {{r^2}''''}{80}}
-{\frac {{r^2}   {f'} f'''  }{48 {f^2}}}
+{\frac {{r^2}  f''''  }{120 f  }}
-{\frac {{r^2}   {f''} ^{2}}{48 {f^2}}}
-{\frac {7{r^2}   {f'}^4}{320 {f^4}}}
-{\frac {7 {{f'}^2}{{(r^2)}''}  }{960 {f^2}}}
+{\frac { {{(r^2)}''} f''  }{120 f  }}
+{\frac {3 {{(r^2)}''}  {{{{(r^2)}'}}^2}}{160 {r^4}}}  \nonumber
\ear
\bear
&&  \left.
+{\frac {13{r^2}   {{f'}^2}f''  }{240 {f^3}}} \right\}
{\frac {d^{2} I_0(\mu)}{d{\mu}^{2}}}
+ \left\{  \left[
-{\frac {3 {{f'}^2} {{{{(r^2)}'}}^2}}{16 {f^2}{r^2}  }}
+{\frac {{r^2}   {{f'}^2}f''  }{8 {f^3}}}
-{\frac { {f'}  {{{(r^2)}'}} f''  }{4 {f^2}}}
-{\frac {{r^2}   {f'}^4}{32 {f^4}}}
\right. \right. \nn
&& \left.
-{\frac { {{(r^2)}''}  {f'} {{{(r^2)}'}}  }{2 f  {r^2}  }}
-{\frac { {{{{(r^2)}'}}^4}}{32 {r^6}}}
+{\frac { {{{{(r^2)}'}}^2}f''  }{8 f  {r^2}  }}
-{\frac {{r^2}   {f''} ^{2}}{8 {f^2}}}
+{\frac { {{f'}^3}{{{(r^2)}'}}  }{8 {f^3}}}
+{\frac { {f'}  {{{{(r^2)}'}}^3}}{8 f   {r^4}}}
+{\frac { {{(r^2)}''}  {{{{(r^2)}'}}^2}}{4 {r^4}}}
\right. \nonumber
\ear
\bear
&& \left.
+{\frac { {{f'}^2}{{(r^2)}''}  }{4 {f^2}}}
-{\frac { {{(r^2)}''} ^{2}}{2{r^2}  }}
-{\frac { {{(r^2)}''} f''  }{2f  }} \right] {\xi}^{2}
+ \left[
{\frac {5{r^2}   {f'} f'''  }{48 {f^2}}}
+{\frac {3 {{(r^2)}''}  {f'} {{{(r^2)}'}}  }{16 f  {r^2}  }}
+{\frac { {f'}  {{{(r^2)}'}} f''  }{3 {f^2}}}   \right. \nonumber
\ear
\bear
&&
-{\frac {5 {f'}  {{{{(r^2)}'}}^3}}{96 f   {r^4}}}
+{\frac { {{(r^2)}''} f''  }{24 f  }}
+{\frac { {{{(r^2)}'}} {{(r^2)}'}''  }{24 {r^2}  }}
-{\frac {{r^2}  f''''  }{24 f  }}
-{\frac {3 {{(r^2)}''}  {{{{(r^2)}'}}^2}}{16 {r^4}}}
+{\frac {5 {{{{(r^2)}'}}^4}}{96 {r^6}}}
-{\frac {17 {{f'}^3}{{{(r^2)}'}}  }{96 {f^3}}}   \nn
&&
+{\frac {5 {{(r^2)}''} ^{2}}{24{r^2}  }}
-{\frac {{r^2}''''}{12}}
+{\frac {11{r^2}   {f'}^4}{96 {f^4}}}
-{\frac { {f'''} {{{(r^2)}'}}  }{8f  }}
-{\frac { {f'} {{(r^2)}'}''  }{12f  }}
+{\frac {{r^2}   {f''} ^{2}}{8 {f^2}}}
+{\frac { {{f'}^2} {{{{(r^2)}'}}^2}}{16 {f^2}{r^2}  }}
-{\frac {7{r^2}   {{f'}^2}f''  }{24 {f^3}}}  \nonumber
\ear
\bear
&& \left.
-{\frac { {{{{(r^2)}'}}^2}f''  }{24 f  {r^2}  }} \right] \xi
-{\frac {49 {f'}  {{{(r^2)}'}} f''  }{960 {f^2}}}
+{\frac { {{{{(r^2)}'}}^2}f''  }{480 f  {r^2}  }}
-{\frac { {{(r^2)}''}  {f'} {{{(r^2)}'}}  }{60 f  {r^2}  }}
+{\frac { {f'}  {{{{(r^2)}'}}^3}}{192 f   {r^4}}}
-{\frac {3 {{(r^2)}''} ^{2}}{160{r^2}  }}  \nn
&&
-{\frac {11 {{{{(r^2)}'}}^4}}{1920 {r^6}}}
+{\frac {13 {{f'}^3}{{{(r^2)}'}}  }{480 {f^3}}}
-{\frac {3 {{f'}^2} {{{{(r^2)}'}}^2}}{640 {f^2}{r^2}  }}
+{\frac {11 {f'''} {{{(r^2)}'}}  }{480 f  }}
+{\frac { {f'} {{(r^2)}'}''  }{80 f  }}
-{\frac { {{{(r^2)}'}} {{(r^2)}'}''  }{240 {r^2}  }}
 \nonumber
\ear
\bear
&&
+{\frac {{r^2}''''}{80}}
-{\frac {{r^2}   {f'} f'''  }{48 {f^2}}}
+{\frac {r^2 f''''}{120 f}}
-{\frac {{r^2}   {f''} ^{2}}{48 {f^2}}}
-{\frac {7 {r^2}   {f'}^4}{320 {f^4}}}
-{\frac {7 {{f'}^2}{{(r^2)}''}  }{960 {f^2}}}
+{\frac { {{(r^2)}''} f''  }{120 f  }} \nn &&
\left.
+{\frac {3 {{(r^2)}''}  {{{{(r^2)}'}}^2}}{160 {r^4}}}
+{\frac {13 {r^2}   {{f'}^2}f''  }{240 {f^3}}} \right\}
\frac {d I_0(\mu)}{\mu d\mu}
+ \left\{  \left[
{\frac { {f'}  {{{(r^2)}'}} f''  }{4 {f^2}}}
-{\frac { {{{{(r^2)}'}}^2}f''  }{8f  {r^2}  }}
+{\frac { {{{{(r^2)}'}}^4}}{32 {r^6}}}
\right. \right. \nn &&
+{\frac {3 {{f'}^2} {{{{(r^2)}'}}^2}}{16 {f^2}{r^2}  }}
+{\frac { {{(r^2)}''} ^{2}}{2{r^2}  }}
+{\frac {{r^2}   {f''} ^{2}}{8 {f^2}}}
+{\frac {{r^2}   {f'}^4}{32 {f^4}}}
-{\frac { {f'}  {{{{(r^2)}'}}^3}}{8 f   {r^4}}}
-{\frac {{r^2}   {{f'}^2}f''  }{8 {f^3}}}
-{\frac { {{(r^2)}''}  {{{{(r^2)}'}}^2}}{4 {r^4}}}
\nonumber
\ear
\bear
&& \left.
-{\frac { {{f'}^3}{{{(r^2)}'}}  }{8 {f^3}}}
+{\frac { {{(r^2)}''} f''  }{2 f  }}
-{\frac { {{f'}^2}{{(r^2)}''}  }{4 {f^2}}}
+{\frac { {{(r^2)}''}  {f'} {{{(r^2)}'}}  }{2 f  {r^2}  }} \right] {\xi}^{2}
+ \left[
-{\frac {5 {{(r^2)}''} ^{2}}{24 {r^2}  }}
-{\frac {5 {{{{(r^2)}'}}^4}}{96 {r^6}}}
\right. \nn &&
+\frac {{r^2}''''}{12}
-{\frac {{r^2}   {f''} ^{2}}{8 {f^2}}}
+{\frac { {{{{(r^2)}'}}^2}f''  }{24 f  {r^2}  }}
+{\frac {7 {r^2}   {{f'}^2}f''  }{24 {f^3}}}
-{\frac { {{f'}^2} {{{{(r^2)}'}}^2}}{16 {f^2}{r^2}  }}
+{\frac {5 {f'}  {{{{(r^2)}'}}^3}}{96 f   {r^4}}}
-{\frac {5 {r^2}   {f'} f'''  }{48 {f^2}}} \nonumber
\ear
\bear
&&
-{\frac { {f'}  {{{(r^2)}'}} f''  }{3 {f^2}}}
+{\frac { {f'''} {{{(r^2)}'}}  }{8 f  }}
+{\frac {3 {{(r^2)}''}  {{{{(r^2)}'}}^2}}{16 {r^4}}}
+{\frac {17 {{f'}^3}{{{(r^2)}'}}  }{96 {f^3}}}
-{\frac {11 {r^2}   {f'}^4}{96 {f^4}}}
-{\frac { {{(r^2)}''} f''  }{24 f  }}
\nn &&
-{\frac {3 {{(r^2)}''}  {f'} {{{(r^2)}'}}  }{16 f  {r^2}  }}
+{\frac { {f'} {{(r^2)}'}''  }{12 f  }}
-{\frac { {{{(r^2)}'}} {{(r^2)}'}''  }{24 {r^2}  }}
\left. +{\frac {{r^2}  f''''  }{24 f  }} \right] \xi
-{\frac {{r^2}''''}{80}}
+{\frac {3 {{(r^2)}''} ^{2}}{160 {r^2}  }}
 \nonumber
\ear
\bear
&&
+{\frac {11 {{{{(r^2)}'}}^4}}{1920 {r^6}}}
-{\frac {13 {{f'}^3}{{{(r^2)}'}}  }{480 {f^3}}}
-{\frac {11 {f'''} {{{(r^2)}'}}  }{480 f  }}
-{\frac { {f'} {{(r^2)}'}''  }{80 f  }}
+{\frac { {{{(r^2)}'}} {{(r^2)}'}''  }{240 {r^2}  }}
-{\frac {{r^2}  f''''  }{120 f  }}
+{\frac {{r^2}   {f''} ^{2}}{48 {f^2}}}
\nn
&&
+{\frac {7{r^2}   {f'}^4}{320 {f^4}}}
+{\frac {7 {{f'}^2}{{(r^2)}''}  }{960 {f^2}}}
-{\frac { {{(r^2)}''} f''  }{120 f  }}
-{\frac {3 {{(r^2)}''}  {{{{(r^2)}'}}^2}}{160 {r^4}}}
+{\frac {49 {f'}  {{{(r^2)}'}} f''  }{960 {f^2}}}
-{\frac { {{{{(r^2)}'}}^2}f''  }{480 f  {r^2}  }}
\nn &&
-{\frac { {f'}  {{{{(r^2)}'}}^3}}{192 f   {r^4}}}
+{\frac {3 {{f'}^2} {{{{(r^2)}'}}^2}}{640 {f^2}{r^2}  }}
+{\frac {{r^2}   {f'} f'''  }{48 {f^2}}}
-{\frac {13 {r^2}   {{f'}^2}f''  }{240 {f^3}}}
\left. +{\frac { {{(r^2)}''}  {f'} {{{(r^2)}'}}  }{60 f  {r^2}  }} \right\} \frac{1}{\mu^2}. \nonumber
\ear

\end{document}